\documentclass[aps,prl,twocolumn,showpacs,superscriptaddress]{revtex4-1}  
\usepackage{amsmath}
\usepackage{amssymb}
\usepackage{graphicx}
\usepackage{epstopdf}
\usepackage{gensymb}
\usepackage[dvipsnames]{xcolor}
\usepackage{multirow}
\usepackage{hhline}

\renewcommand{\vec}{\mathbf}

\definecolor{HH-color}{named}{magenta}
\definecolor{HH-color2}{named}{NavyBlue}

\begin{document}

\setlength{\abovedisplayskip}{3.5pt}
\setlength{\belowdisplayskip}{3.5pt}

\preprint{APS/123-QED}

\title{Charged Defects and Phonon Hall Effects in Ionic Crystals}
% Force line breaks with \\
%\thanks{A footnote to the article title}%

\author{B. Flebus}
\affiliation{Department of Physics, Boston College, 140 Commonwealth Avenue Chestnut Hill, MA 02467}
\author{A.H. MacDonald}
\affiliation{Physics Department, University of Texas at Austin, Austin TX 78712}

\date{\today}% It is always \today, today,
             %  but any date may be explicitly specified

\begin{abstract}
It has been known for decades that a magnetic field can deflect phonons as they flow in response to a thermal gradient,
producing a thermal Hall effect.  Several recent experiments have revealed 
ratios of the phonon Hall conductivity $\kappa_H$ to the phonon longitudinal conductivity $\kappa_L$ 
in oxide dielectrics that are larger than $10^{-3}$ when phonon mean-free-paths exceed phonon wavelengths.
At the same time $\kappa_H/\kappa_L$ is not strongly temperature dependent.
We argue that these two properties together imply a mechanism related to phonon scattering 
from defects that break time-reversal symmetry, and we show that 
Lorentz forces acting on charged defects produce substantial
skew-scattering amplitudes, and related thermal Hall effects that are consistent with recent observations.  
%We comment on a possible interpretation of the sudden change 
%observed in the size of the thermal Hall effect as a function of doping in cuprates based on this mechanism.

\end{abstract}

%A thermal Hall response can originate from coupling of phonons to spins or other magnetic-field 
%sensitive electronic degrees of freedom that imprints chirality on the phonon system.
%This mechanisms, however, yields a thermal Hall conductivity $\kappa_{H}$ 
%that is negligible compared with its longitudinal counterpart $\kappa_{L}$, and decreases more 
%rapidly at low-temperatures.

%\keywords{Suggested keywords}%Use showkeys class option if keyword
                              %display desired
\maketitle

%\tableofcontents

%\begin{itemize}
%         \item   Formal wave equation for isotropic elastic media including a Lorentz force  
%     \item   Formal expression for the scattering amplitude in terms of T-matrix and discussion of correction at first
%     order in the Lorentz force.
%     \item   Brief summary of the solution in the absence of the Lorentz force.  Fit to experimental data for LSCO? 
%     Plot of the scattering amplitude squared as a function of scattering angle for some typical parameters? 
%     Discussion of incoming and outgoing state representations.
%   The outgoing solutions are just their time-reversed counterparts - which are their complex conjugates 
%         \item  Discussion of recent experiments 
%\end{itemize}

\textit{Introduction}---
In recent years the  thermal Hall effect has frequently been employed as an informative probe of strongly correlated  materials~\cite{Griss2019,Boulanger2020,GrissNP2020,Li2020,Kasahara2018,Zhang2010,Zhang2011,Mori2014,SachdevPRB2019,Sachdev2019,Lee2019,Kivelson2020,Lee2015,Nasu2017,Hentrich2019,Kasahara2018a,Hirschberger2015,Ye2021,Guo2021}.  
In the process, it has become clear that relatively large thermal Hall conductivities ($\kappa_H$) that are 
linearly proportional to a magnetic field $B$ are common in oxide dielectrics.
The linear dependence of $\kappa_H$ on $B$ is expected since this non-reciprocal transport coefficient requires 
time-reversal symmetry breaking.  What is surprising is not that $\kappa_H/B \ne 0$, but that it is relatively large.  

Large thermal Hall conductivities are not limited to magnetic materials,
and even in magnetic materials usually have an onset that is not related to the onset of magnetic order~\cite{Griss2019}.
In $\text{La}_2 \text{CuO}_4$ the thermal Hall conductivity is almost isotropic~\cite{GrissNP2020},
like the phonon spectrum, whereas the magnon spectrum is quasi-two-dimensional.
The dominant source of the thermal Hall effect is therefore not magnon transport, at least in most cases,
although there must be a magnon Hall effect and its theoretical description is certainly interesting~\cite{Onose2010,Ideue2012,Murakami2011,Murakami20112, Greg2017}.
Phonons, the dominant heat carriers in most dielectrics, evidently have a Hall effect~\cite{Kivelson2020}.
For magnetic fields $\sim 10$ Tesla, the ratio of $\kappa_H$ to 
the longitudinal thermal conductivity $\kappa_L$ is often larger than $10^{-3}$ over a wide range of temperatures~\cite{Griss2019,Boulanger2020,GrissNP2020,Li2020}.
 
\begin{figure}[b]
\centering
\includegraphics[width=0.84\linewidth]{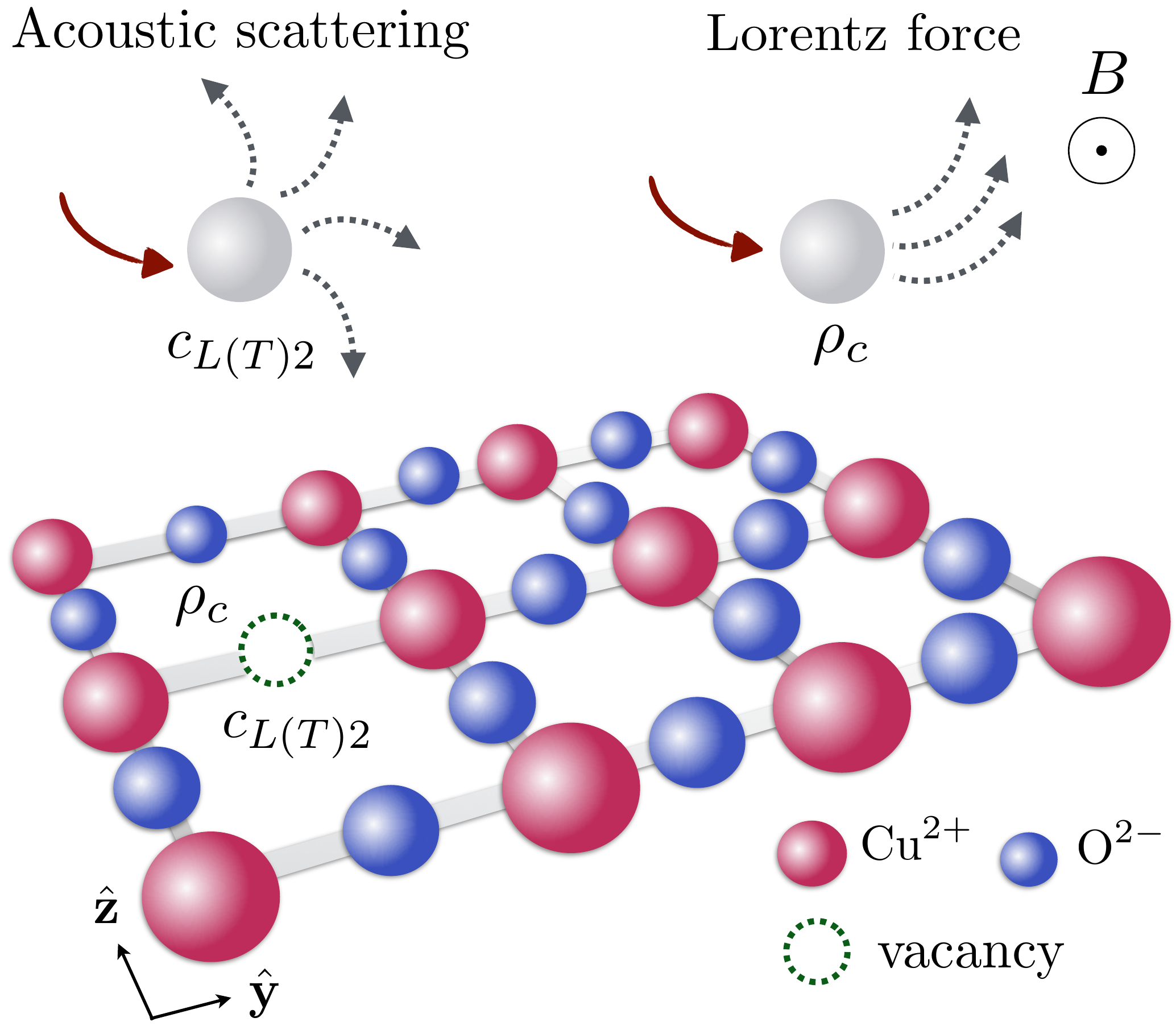}
\caption{Oxygen vacancies, like the one depicted in this two-dimensional cartoon, are common defects in oxide dielectrics. 
We model them as spherical scatterers that have a finite net charge density $\rho_c$, and 
longitudinal, $c_{L2}$, and transverse, $c_{T2}$, local sound velocities that differ from those of the surrounding medium.
In the presence of an out-of-plane magnetic field $B$, 
acoustic waves will experience a Lorentz force inside the defect region that combines 
with normal acoustic scattering to add a small non-reciprocal 
contribution to the phonon scattering rates.}
\label{Fig1}
\end{figure}

Large thermal Hall effects are normally observed in a temperature range over which the 
the phonon mean-free-path $\ell$ exceeds the wavelengths of thermally active phonons, $\lambda_T \sim \hbar c/k_B T$,
where $c$ is the mode velocity.  
When this condition is satisfied, phonon transport can be described using a Boltzmann equation, 
and the phonon conductivity tensor is limited by phonon scattering.
The non-reciprocity could in principle originate from an intrinsic mechanism that 
acts between scattering events, or from a non-reciprocal property of the extrinsic scatterers.
Mechanisms responsible for intrinsic chirality in phonon transport have been 
extensively investigated~\cite{Kivelson2020,Mori2014,Zhang2010,Zhang2011,Xiao2019, 1PhononHall}.
For example, coupling to a spin environment can provide phonon bands with   
a finite Hall viscosity, $\eta_H$, \cite{Ye2021,Guo2021} which characterizes the strength of 
the time-reversal symmetry breaking inherited from the spin system.
Magnetic fields also influence the lattice dynamics of ionic crystals 
directly through the Lorentz forces that act on moving ions~\cite{Flebus2021}.  
The Lorentz force couples longitudinal in-phase motion of the cations and anions
to out-of-phase transverse motion.  At small $k$ the chiral component of 
acoustic phonon polarization vectors vanishes like $(ka)^2$.  It follows that $\eta_H$ 
has a Lorentz force contribution $\sim Z^* e B/(c a) = M \Omega_{cI}/a$, where $\omega_{cI} = Z^* eB/Mc $ is the 
ion cyclotron frequency, $Z^*$ is the effective ion charge, 
$M$ is the ion mass and $a$ is the crystal lattice constant.  
Whatever its origin, Hall viscosity combined with non-chiral phonon scattering always produces a thermal 
Hall conductivity.  Because the chirality of the active phonons is proportional to $k^2$, and 
the typical $k \sim \lambda_T^{-1} \propto T$, this mechanism always yields a $\kappa_H/\kappa_L$ ratio that 
declines with decreasing temperature.  (Guo {\it et al.} have recently concluded that the intrinsic mechanism yields 
$\kappa_{H}/\kappa_{L} \sim T^4$ \cite{Guo2021} behavior.)
Experimentally $\kappa_H/\kappa_L$ is nearly temperature independent, except sometimes
at the lowest temperatures where boundary scattering starts to play a role, pointing to an extrinsic mechanism. 
In order to retain its impact at low temperatures, time-reversal-breaking must be 
embedded in the properties of the phonon scatterers, not the phonon medium. 

Our model for phonon Hall effects applies when two conditions are satisfied: i) we require that the phonon mean free path $\ell$  exceeds the thermal wavelength $\lambda_{T}$ of acoustic phonons with energy $\hbar \omega \sim k_B T$, i.e., $\ell \gg \lambda_T$. The mean free path can be estimated from measured thermal conductivities and heat capacities using the relationship  $\ell \sim \kappa/cC$, where $C$ is the heat capacity. The thermal wavelength is of order $\sim a T_D/T$, where $T_D$ is the acoustic phonon Debye temperature, and 
$a$ is the crystal lattice constant. 
 It has been understood for decades \cite{Pomeranchuk,Klemens1951} that 
phonon transport in this regime, which is applicable in good crystals, can be described using a Boltzmann equation.
ii) The ratio of the Hall conductivity to the longitudinal conductivity does not 
decrease monotonically with temperature even well below the Debye temperature, as it would if an 
extrinsic Hall viscosity mechanism were dominant.  
In the temperature range of interest, scattering of long-wavelength longitudinal acoustic (LA) phonons by 
defects limits thermal transport~\cite{Pomeranchuk}. 
Oxygen vacancies and interstitials  are  prevalent defects in many oxide dielectrics. The removal or the addition of an atom from a lattice site disrupts the local bonding, yielding strain fields that produce a crudely spherical defect with elastic 
properties that differ from those of the surrounding medium, as depicted  in Fig.~\ref{Fig1}.
Since oxygen vacancies and interstitials are dopants, the ions in the neighborhood 
of the defect have a net charge, and will therefore experience a Lorentz force.
In this Letter we propose that the thermal Hall effect in many oxide dielectrics is 
due to non-reciprocal phonon scattering from charged defects. Specifically, we show that a contribution to phonon skew scattering that is linear in magnetic field results from interference between Lorentz force and 
acoustic potential defect scattering.

\textit{Phonon Boltzmann Equation}--- We consider the steady-state phonon distribution function
$f_{\vec{k},s}$ of a system driven from equilibrium by a temperature gradient $\nabla T$.
(Here $\vec{k}$ is the phonon wavevector and $s$ the mode label.)
We write $f_{\vec{k},s}=f^{(0)}_{\vec{k},s}+g_{\vec{k},s}$, where  $f^{(0)}_{\vec{k},s}$  
is the equilibrium Bose-Einstein distribution function and $g_{\vec{k}s}$ is linear in $\nabla T$. 
At low temperatures the thermal conductivity is limited by elastic scattering 
of longitudinal phonons~\cite{Klemens1951,Pomeranchuk}.
The Boltzmann equation therefore reads
\begin{widetext}
\begin{equation}
 \omega_{\textbf{k}} \; \frac{\partial \omega_{\textbf{k}}}{\partial \vec{k}} \cdot \left( \frac{\nabla T}{T} \right) \left( - \frac{\partial f^{(0)}_{\vec{k}}}{\partial \omega_{\textbf{k}}} \right)  = \sum_{\mathbf{q}} \left( W_{\vec{q} \rightarrow \vec{k}} \; g_{\vec{q}}- W_{\vec{k} \rightarrow \vec{q}} \; g_{\vec{k}} \right)\,,
 \label{eq:Boltzmann}
\end{equation}
\end{widetext}
where we have restricted attention to the longitudinal phonon mode 
with frequency $\omega_{\textbf{k}} $. Here, $W_{\vec{k} \rightarrow \vec{q}}$ 
is the rate of elastic scattering from initial state $(\vec{k})$ to final state 
$(\vec{q})$.
For temperatures well below the Debye temperature,  the phonon Hall conductivity 
is largest, $\omega_{\textbf{k}} \approx c |\textbf{k}|$, where $c$ is the longitudinal phonon velocity.
Non-reciprocal (skew) phonon scattering ($W_{\vec{q} \rightarrow \vec{k}} \ne  W_{\vec{k} \rightarrow \vec{q}}$)
requires broken time-reversal symmetry that is, in the case of interest, supplied by an external magnetic field
$\vec{B}=B \hat{\vec{n}}_{B}$, where $\hat{\vec{n}}_{B}$ is a unit vector along the direction of the field.  
For a cubic crystal with short-range scatterers, 
 \begin{align}
W_{\vec{k} \rightarrow \vec{q}}= \frac{1}{V} \tau_{\omega}^{-1} \nu_{\omega}^{-1} \left[ 1 - \Lambda_{\omega} \hat{\vec{n}}_{B} \cdot \left( \hat{\mathbf{k}} \times \hat{\mathbf{q}} \right) \right]\, \delta(\omega-cq),
\label{2311}
\end{align} 
where $V$ is the system volume, $\tau_{\omega}$ is the phonon relaxation time, $\nu_{\omega}$ is the phonon density of states,  
and  $\Lambda_{\omega} \ll 1$  is a 
small parameter, calculated explicitly below, that
characterizes the ratio of nonreciprocal to reciprocal phonon scattering. 
When inserted in the Boltzmann equation, Eq.~(\ref{2311}) yields the longitudinal ($\kappa_{L}$)
and Hall ($\kappa_{H}$) thermal conductivities:
\begin{align}
\kappa_{L}&=\frac{k_B^4 T^3}{2\pi^2 \hbar^3 c} \int dx \; \tau_{\omega}  \frac{ x^4 e^x}{(e^x-1)^2}\,, \label{longitudinalheat} \\
\kappa_{H}&=\frac{k_B^4 T^3}{2\pi^2 \hbar^3 c} \int dx \; \frac{\Lambda_{\omega}\tau_{\omega} }{3}  \frac{x^4 e^x}{(e^x-1)^2}\,,
\label{heatconductivity}
\end{align} 
with  $\hbar\omega = k_B T x$.  
It follows that $\kappa_{H}/\kappa_{L} \sim \Lambda_{\omega}$ at $\omega \sim k_B T$.  

\textit{Low-Temperature Phonon Scattering} ---
Because phonon scattering from bulk defects declines when phonon wavelengths 
exceed defect sizes, the thermal phonon mean-free path $\ell =c \tau_{\omega}
 \propto T^{-4}$ \cite{Pomeranchuk,Klemens1951},
and is limited only by the system size $L$ in the low temperature limit.
Experiments show that the Hall conductivity is large and weakly temperature dependent when 
$L \gg \ell \gg \lambda$.  We argue that this behavior can be explained by scattering from defects, 
for example oxygen vacancies or interstitials, that act as dopants.
Since the ions in the vicinity of a dopant complex have a net charge,
the local Lorentz force does not vanish in the interior of the scattering center, 
as sketched in Fig.~\ref{Fig1}.
Below we show that a contribution to phonon skew scattering that is linear in magnetic field  
results from the interference between this Lorentz force and the acoustic scattering potential.  
The strength of the effect can be characterized by the ion-cyclotron frequency
$\omega_c$, which is $\sim 10^5$ Hz at $B=10$ T, depending on the ion charge and mass.

In order to obtain an  explicit form for the scattering amplitude 
we first examine acoustic scattering in the absence of a magnetic field. 
It is convenient to rewrite the acoustic wave equation in this limit as 
\begin{align}
\left( \nabla^2 + K_{n}^2 \right) \vec{u}_{n}(\mathbf{r}) - \left(1-\frac{K_{n}^2}{k_{n}^2}  \right) \nabla \left[\nabla \cdot \vec{u}_{n}(\mathbf{r}) \right]=0\,,
\label{1877}
\end{align}
where   $K_{n}=\omega/c_{Tn}$  and $k_{n}=\omega/c_{L n}$, where $n=1(2)$ labels the region outside (inside) the defect.   Here  
$c^2_{Tn}=\mu_{n}/\rho_{n}$ and $c^2_{Ln}=(\lambda_{n}+2\mu_{n})/\rho_{n}$ are, respectively, 
the squares of the transverse and longitudinal phonon velocities,  
$\rho_{n}$ is the mass density, and $\lambda_{n}$ and $\mu_{n}$ are Lam\'e constants. 
The small absorption losses that are always present can be accounted for~\cite{Ayres} 
by modeling both the homogeneous and the defect region as a fluid with kinetic viscosity $\eta_{1(2)} \omega \ll \lambda_{1(2)},\mu_{1(2)}$,
by letting 
\begin{eqnarray}
\mu_{1(2)} &\rightarrow& \mu_{1(2)} - i\eta_{1(2)} \omega\,,\nonumber \\
\lambda_{1(2)} &\rightarrow& \lambda_{1(2)} - i\eta_{1(2)} \omega\,.
\end{eqnarray}
In the following,
 we take the $z$-direction as the direction of propagation of the longitudinal incident wave, i.e., $\mathbf{k}=k\hat{\mathbf{z}}$~\cite{note}.
When a longitudinal wave impinges on the 
surface of the spherical scatterer, it can be scattered either as a longitudinal 
wave or as a transverse wave.  The scattering problem (\ref{1877})
can be  conveniently solved in spherical coordinates 
by introducing the scalar functions $\pi_{Ln}$ and $\pi_{Tn}$ defined by  ~\cite{hinders,supp}
\begin{eqnarray}
\mathbf{u}_{L n}&=&-\frac{1}{k_{n}^2}\nabla \pi_{Ln}\,, \; \; \; \; \nonumber \\
\mathbf{u}_{T n}&=&\frac{1}{K_{n}} \nabla \times \nabla \times (\mathbf{r} \pi_{Tn})\,.
\end{eqnarray}
The scalar potentials corresponding to incident ($^i$), transmitted ($^t$), and scattered ($^s$) waves can be written as~\cite{hinders}
\begin{align}
r\pi^{i}_{L}=&\sum^{\infty}_{l=0} i^{l+1} (2l+1) \psi_{l}(k_{1} r) P_{l}(\cos \theta)\,,  \\
r\pi^{t}_{L}=&\sum^{\infty}_{l=0}  i^{l+1} A_{l} (2l+1) \psi_{l}(k_{2} r) P_{l}(\cos \theta)\,, \\
r\pi^{t}_{T}=&\frac{1}{K_{2}}\sum^{\infty}_{l=1}  i^{l+1} B_{l} (2l+1)  \psi_{l}(K_{2} r) P_{l}(\cos \theta)\,, \\
r\pi^{s}_{L}=&\sum^{\infty}_{l=0} i^{l+1} C_{l} (2l+1)   \zeta_{l}(k_{1} r) P_{l}(\cos \theta)\,,  \\
r\pi^{s}_{T}=&\frac{1}{K_{1}}\sum^{\infty}_{l=1}  i^{l+1}D_{l}  (2l+1)  \zeta_{l}(K_{1} r) P_{l}(\cos \theta) \,,
\label{228}
\end{align}
with
\begin{align}
&\psi_{l}(x)=\sqrt{x\pi/2} J_{l+1/2}(x), \nonumber \\
& \zeta_{l}(x)=\sqrt{x\pi/2} H^{(1)}_{l+1/2}(x)\,,
\end{align}
where $J_{l+1/2}(x)$ and $H^{(1)}_{l}(x)$ are the half-order cylindrical Bessel and Hankel functions, $P_{l}(\cos \theta)$ is the Legendre function of degree $l$, and $\theta$ is the scattering angle. The coefficients $A_{l}$, $B_{l}, C_{l}$ and $D_{l}$ are obtained by imposing  
the continuity of the displacement field and the stress tensor at the boundary radius $r=a$:
\begin{equation}
u^{t}_{r(\theta)}=u^{i}_{r (\theta)}+u^{s}_{r (\theta)}\,, \; \; \; \; \; \; \sigma^{t}_{rr(r\theta)}=\sigma^{i}_{rr(r\theta)}+\sigma^{s}_{rr(r\theta)}\,,
\label{1911}
\end{equation}
where the stress tensor components 
\begin{align}
&\sigma_{rr}=\lambda \nabla \cdot \mathbf{u} + 2 \mu \partial_{r} u_{r}\,,  \\
&\sigma_{r\theta}=\mu \left( \partial_{r} u_{\theta} -\frac{u_{\theta}}{r} + \frac{1}{r} \partial_{\theta} u_{r} \right)\,.
\end{align}
If the difference in mass density and Lam\'e constants 
between the two regions is small, i.e., $\rho_{1}/\rho_{2}, \mu_{1}/\mu_{2}, \lambda_{1}/\lambda_{2} \sim 1$, one can safely neglect wave interconversion, i.e., $D_{l} \ll C_{l}$ for each $l$. Thus, most of the incident longitudinal acoustic wave amplitude will be scattered as a longitudinal wave. 
In the Rayleigh scattering regime~\cite{Rayleigh}, the wavelength of acoustic waves is much larger than the defect size, i.e., $ka\ll 1$. 
To leading order in $ka$ the scattered wave is dominated by its $l=0$ component 
\begin{align}
\mathbf{u}_{\mathbf{k}'L}(\mathbf{r})=\hat{\mathbf{e}}_{L}  \frac{e^{ikr}}{r}f_{\mathbf{k} L \rightarrow \mathbf{k}' L}\,,
\end{align}
where $\hat{\mathbf{e}}_{L}=\left( \cos \phi \sin \theta,  \sin \phi \sin \theta, \cos \theta \right)$ and $\mathbf{k}'=k\hat{\mathbf{e}}_{L}$ are, respectively,  the polarization and the wavevector of the outgoing LA wave. Here $f_{\mathbf{k} L\rightarrow \mathbf{k}' L}$ is the longitudinal 
scattering amplitude from $\mathbf{k}$ to $\mathbf{k}'$:
\begin{align}
f_{\mathbf{k} L\rightarrow \mathbf{k}' L}= \frac{\text{Im} C_{0} - i \text{Re} C_{0}}{k}\,,
\label{scattampl}
\end{align}
with $\left(\text{Im}C_{0}\right)^2+\left( \text{Re}C_{0}\right)^2=-\text{Re}C_{0}$, as dictated by the extinction theorem~\cite{optics}. Retaining only terms linear in the small parameters $\eta_{1(2)} \omega/\mu_{1(2)}$, we find
\begin{align}
\text{Im} C_{0} \simeq&  +i \frac{(ka)^3}{3} \frac{3\left( \lambda_{1}-\lambda_{2} \right) + 2 \left( \mu_{1}-\mu_{2} \right)}{3\lambda_{2}+2 \left(2\mu_{1}+\mu_{2} \right)}  \,, \label{263}  \\
\text{Re}C_{0}\simeq&  +\frac{\eta_{1} \omega \left(  4 \mu_{1} + 6\mu_{2} + 9\lambda_{2}-4\lambda_{1} \right)  }{\left( 3\lambda_{2}+ 4\mu_{1}+2\mu_{2} \right)^2}   (ka)^3 \nonumber \\
& - \frac{5\eta_{2} \omega \left( \lambda_{1} + 2\mu_{1} \right)}{\left( 3\lambda_{2}+ 4\mu_{1}+2\mu_{2} \right)^2} (ka)^3 \,. \label{264}
\end{align}

\textit{Interference between Rayleigh and Lorentz Scattering}---
Working in the continuum elasticity theory limit valid at low temperatures, 
we describe the phonons by a displacement vector field $\mathbf{u}(\mathbf{r})$ 
that satisfies the following wave equation:
\begin{equation}
\sum_{j=x,y,z} \Big[{A}^{0}_{ij} + A^{R}_{ij} + A^{L}_{ij} - \omega^2 \delta_{ij} \big] u_j(\vec{r}) =0\,,
\label{fullWE}
\end{equation}
where $A^{0}_{ij}$ is the acoustic differential operator in the absence of a magnetic field, 
$A^{R}_{ij}$ is the difference between the acoustic differential operator inside and outside the defect region,
accounted for in the previous section, and $A^{L}$ is a Lorentz force term that acts only inside the defect. 
For a uniform isotropic medium 
\begin{align}
A^{0}= 
\begin{pmatrix} c_{1}^2 k_x^2 + c^2_{T1} k^2 &  c_{1}^2 k_x k_y  & c_{L1}^2 k_{x} k_{z} \\   c_{1}^2 k_{x} k_{y} &  c^2_{T1} k^2 + c_{1}^2 k^2_{y} & c_{1}^2 k_{y} k_{z} \\ c_{1}^2 k_{x} k_{z}  & c_{1}^2 k_{y} k_{z} & c_{1}^2 k^2_{z} + c^2_{T1} k^2 \end{pmatrix}\,,
\label{dynamicalmatrix}
\end{align}
where $k_{i} = -i \nabla_i$, $k^2=k_x^2+k_y^2+k_z^2$, and $c_{1}^2=c^2_{L1}-c^2_{T1}$.  
For a given frequency $\omega$, there are three independent solutions 
unperturbed elastic waves with vector displacement fields $\vec{u}_{\mathbf{k} \alpha}(\vec{r})= \hat{\mathbf{e}}_{\alpha}  \cos(\vec{k} \cdot \vec{r})$, where $|\vec{k}| = \omega/c_{L1}$ for the longitudinal mode 
($\vec{k} \times \hat{\mathbf{e}}_{L1}=0$) and  $|\vec{k}| =\omega/c_{T1}$ 
for two degenerate transverse ($\vec{k} \cdot \hat{\mathbf{e}}_{\alpha}=0)$ modes. 

%In the temperature regime we focus on, phonon heat transport in a crystal is normally limited by scattering of phonons off atomic-scale defects~\cite{Klemens1951} that can be modeled as spheres of radius $R$ 
%within which the local Lam\'e constants ($\lambda_{2}, \mu_{2}$) and mass density ($\rho_{2}$) differ from medium values.
%At low temperatures the wavelengths of active phonons increase and the Rayleigh scattering 
%limit~\cite{Rayleigh}, in which the defect size is 
%small compared to the phonon wavelength ($kR\ll1$), is approached.  
The ions that surround a charged defect are subject to a Lorentz force that is
perpendicular to the applied magnetic field and to the ion velocity.  Our goal is to 
calculate the corrections to the phonon scattering rate that are linear in Lorentz force, and 
hence in magnetic field. This correction is guaranteed by time-reversal symmetry to be non-reciprocal.  
The Lorentz force contribution to the acoustic differential operator 
\begin{equation}
A^{L}_{ij} = i \omega \omega_c ( \hat{\mathbf{j}}\times \hat{\mathbf{i}} ) \cdot \textbf{n}_{B}\,,
\label{scattL}
\end{equation}
is non-zero inside the defect sphere.  Here $\omega_c = \rho_c B/\rho_{2}$ is  the effective ion cyclotron 
frequency and $\rho_{c}$ is the charge density of the defect region.

The scattering rate from an incoming longitudinal wave with  
wavevector $\vec{k}$ to an outgoing unperturbed wave with wavevector $\vec{k}'$ is related to the acoustic scattering 
T-matrix by
\begin{equation}
W_{\vec{k} \rightarrow \vec{k}' } = \frac{ 2\pi}{ V^2\omega^2 }\;  |\langle \vec{k}' |T|\vec{k}\rangle|^2 \; 
\delta(\omega_{\vec{k'}}-\omega)\,,
\label{rate}
\end{equation}
where $T=(A^R+A^L)+(A^R+A^L)G^{0}T$ is the total acoustic scattering T-matrix and 
\begin{equation}
 G^{0} = \Big[\delta_{ij} \, (\omega+i\eta)^2- A^{0}_{ij} \Big]^{-1}\,
\end{equation}
is the unperturbed acoustic Greens function.  To first order in $A^L$,
$T=T^R+T^R (A^{R})^{-1} A^L (A^R)^{-1} T^R + \ldots$~\cite{textbooks}, where 
$T^R=A^R+A^R G^0 T^R$ is the Rayleigh scattering T-matrix, which is related to the 
scattering amplitude calculated above by
\begin{align}
&f_{\mathbf{k} L \rightarrow \mathbf{k}' L}=-\frac{1}{4\pi} \frac{1}{c^2_{L1}} \langle \vec{k}', L |T^{R}| \vec{k}, L \rangle\,.
\label{scattR}
\end{align}
Because the $B=0$ long-wavelength phonon scattering is weak, in the long-wavelength Rayleigh limit we can approximate 
$T^R (A^{R})^{-1} \approx I$, which yields $T \approx T^R+A^L$ where 
\begin{align}
\langle \mathbf{k}', L |A^{L}| \mathbf{k}, L \rangle &= i \omega \omega_{c} \left( \hat{\mathbf{z}} \times \hat{\mathbf{e}}_{L} \right) \cdot \hat{\mathbf{n}}_{B} \nonumber \\ \times &\int d^3 \mathbf{r} \; e^{-i(\mathbf{k}'-\mathbf{k})\cdot \mathbf{r}} \Theta(a-r) \nonumber \\
&\underbrace{\simeq}_{ka\ll 1} i V_{\text{sp}} \omega \omega_{c}  \left( \hat{\mathbf{z}} \times \hat{\mathbf{e}}_{L} \right) \cdot \hat{\mathbf{n}}_{B}\,,
\label{Lorentzfinal} 
\end{align}
and $\Theta(x)$ is the Heaviside step function and  $V_{\text{sp}}=4\pi a^3/3$ is the  defect volume.

\textit{Phonon Hall Effect}---
Combining Eqs.~\eqref{Lorentzfinal},~\eqref{rate},~\eqref{scattR}, and~\eqref{scattampl}, 
and assuming that the dielectric contains a density $n_{s}$ of randomly distributed charged defects, we obtain the expression 
employed below to estimate the phonon Hall effect: 
\begin{align}
&W_{\vec{k} L \rightarrow \vec{k}' L} =\frac{2\pi n_{s}}{ V \omega^2  }  \delta(\omega_{\vec{k}' L}-\omega) \nonumber \\
&\times \left\lvert 4\pi c^2_{L1}\frac{i \text{Re} C_{0}-\text{Im}C_{0} }{k} +i V_{\text{sp}} \omega \omega_{c} (\hat{\mathbf{z}} \times \hat{\mathbf{e}}_{L})  \cdot \hat{\mathbf{n}}_{B} 
 \right\rvert^2\,.
 \label{W262}
\end{align}
Random Lorentz forces would on their own yield a phonon scattering rate proportional to 
$B^2$, and a skew scattering rate proportional to $B^3$.  The linear in $B$ effect observed experimentally
must therefore arise from interference between Rayleigh and Lorentz scattering terms.
Retaining only the linear terms and setting $\omega=c_{L1}k$, we obtain an expression for the 
dimensionless skewness parameter employed in Eq.~\eqref{heatconductivity}:
\begin{align}
\Lambda_{\omega}\simeq \frac{\omega_{c} \eta_{1} \left(   \frac{ 4 \mu_{1}}{\lambda_{1}} +  \frac{6 \mu_{2}}{\lambda_{1}} + \frac{9\lambda_{2}}{\lambda_{1}}-4 \right) - 5 \omega_c \eta_{2}
\left( 1 + \frac{2\mu_{1}}{\lambda_{1}} \right) }{\lambda_{1} \left[ \left( 1-\frac{\lambda_{2}}{\lambda_{1}} \right) + \frac{2}{3} \left( \frac{\mu_{1}}{\lambda_{1}}-\frac{\mu_{2}}{\lambda_{1}} \right)\right]^2}\,.  
\label{267}
\end{align}
The right-hand-side of Eq.~(\ref{267}) is energy-independent, implying that 
in the Rayleigh scattering regime, the Hall to longitudinal conductivity ratio is temperature independent,
with both quantities $\propto T^{-1}$.  

We estimate the typical values of $\Lambda_{\omega}$ at magnetic field $H=15$ \text{T} and temperature $T=15$ K
by setting $\omega_{c}$ to the oxygen ion cyclotron frequency, with $\rho_{c}>0$ for oxygen vacancies~\cite{Gunkel}, and 
assuming a 1\% difference for the Lam\'e constants inside and outside the defect region with $\lambda_{1}, \mu_{1} > \lambda_{2}, \mu_{2}$, 
and setting  $\eta
_{2} \omega/\lambda_{1} > \eta_{1} \omega/\lambda_{1} \sim 10^{-2}$.  These estimates yield 
\begin{align}
\frac{\kappa_{H}}{\kappa_{L}} \sim - 10^{-3} \,,
\label{276}
\end{align}
which is consistent with the order of magnitude observed in experiment~\cite{Griss2019,Boulanger2020,GrissNP2020}.
Note that the skewness in these estimates is larger than the ratio of the ion cyclotron frequency at 15 T to the thermal phonon frequency at 
$T=15$ K because the elastic constant jump near the defect is assumed to be small in relative terms;  skew scattering is larger in 
relative terms because the reciprocal scattering processes are weak.
The sign of the thermal Hall conductivity is negative in many systems - also in agreement with our result~(\ref{276}). 
It must be noted, however, that the sign of Eq.~(\ref{276}), as well as its magnitude,  
is very sensitive to  the relative strength of elastic and attenuation constants. Both positive and negative signs 
are possible in our interpretation.  When elastic scattering from electrically neutral defects 
plays a more important role, the $\kappa_{H}/\kappa_{L}$ (\ref{276}) ratio should decline.
The emergence of an important role for boundary scattering, which normally dominates in the low-temperature limit,
is signaled experimentally by a maximum in $\kappa_L(T)$ at a finite temperature $T_{max}$.  The explanation for the 
phonon Hall effect predicts, in agreement with experiment, that $\kappa_{H}/\kappa_{L}$ begins to decrease rapidly for $ T \ll T_{max}$.

\textit{Discussion}---
Large thermal Hall conductivity signals have been observed in high-temperature superconductors 
over a wide range of doping between insulating and over-doped states~\cite{Griss2019,Boulanger2020,GrissNP2020}.
Since phonon chirality is observed to change continously, decreasing gradually with increases in doping, it is natural
to assume that the same mechanism applies in insulating and pseudogap states.  If the phonon Hall effect is 
indeed due to scattering from charged defects, these would have to retain a local effective charge in the pseudogap 
state.  That is to say that local screening by mobile electronic quasiparticles would have to be imperfect at the thermal phonon time scale,
on the length scale of the defect.  In cuprates phonon chirality drops upon exiting the pseudogap state, which is 
consistent with strengthening screening.  Note that the Lorentz force on an ion in a doped ionic crystal vanishes only if ionic charges are perfectly screened locally, a condition that is approached only in good metals.

In summary, we have constructed a model of thermal transport by chiral phonons.  The phonon conductivity is 
limited by scattering from charged defects.  In our model long-wavelength acoustic phonons experience 
both elastic and Lorentz force, due to the unscreened charge of the bound dopants.
We have shown that the puzzling giant thermal Hall signal observed recently in many dielectric oxides can be 
explained by the interference between elastic and Lorentz acoustic potentials. 
The estimated magnitude and sign of the effect is consistent with the low-temperature experimental 
observations~\cite{Griss2019,Boulanger2020,GrissNP2020}. 
Kinetic viscosity in the crystal is required to get a thermal Hall effect that is linear in field. 
Future studies should address more systematically the parameters describing the  elastic properties of oxygen vacancies as well as their 
effective charge in insulating and pseudogap phases.  We hope that our work will stimulate further experimental 
investigations of the role of charged defects in phonon-driven thermal transport. 
\\

\textit{Acknowledgments.} The authors thank  S. Kivelson, M. E. Boulanger, G. Grissonnanche, B. Ramshaw, and L. Taillefer  for helpful discussions. 
This work was supported by ONR MURI N00014-1-1-2377.

\appendix

% The \nocite command causes all entries in a bibliography to be printed out
% whether or not they are actually referenced in the text. This is appropriate
% for the sample file to show the different styles of references, but authors
% most likely will not want to use it.
%\nocite{*}
%V. M. Ayres and G. C. Gaunaurd, J. Acoust. Soc. Am. 81, 301
% 1987 .

%Lee2015,Nasu2017,Hentrich2019,Kasahara2018,Hirschberger2015

%\bibliography{}% Produces the bibliography via BibTeX.

\begin{thebibliography}{99}
\bibitem{note} Note that this is an arbitrary choice and in principle every direction should be accounted for.
\bibitem{supp} There is an additional transverse (horizontal shear) wave solution. However, the horizontal shear components are not included in our discussion because they are not excited by an incident compressional wave.
\end{thebibliography}

\end{document}